\documentclass{elsart}

\usepackage{epsfig}

\usepackage{amssymb}

\begin{document}
\begin{frontmatter}




\title{Temperature dependence of the upper critical field of high-$T_{c}$ 
superconductors from isothermal magnetization data. Application to 
polycrystalline samples and ceramics.}

\author[label1,label2]{I. L. Landau}
\author[label1]{H. R. Ott}
 \address[label1]{Laboratorium f\"{u}r Festk\"{o}rperphysik, 
 ETH-H\"{o}nggerberg, CH-8093 Z\"{u}rich, Switzerland}
 \address[label2]{Institute for Physical Problems, 117334 Moscow, Russia}



\begin{abstract}
	
Using a recently developed scaling procedure that allows to establish the 
temperature  dependence of the normalized upper critical field 
$h_{c2}(T) = H_{c2}(T)/H_{c2}(T_{0})$ from measurements of the equilibrium 
isothermal magnetization, we analyze experimental data obtained on 
polycrystalline and ceramic samples. We show that the scaling procedure 
works quite well in all cases that we have considered. This provides a very 
strong evidence that the temperature dependencies of $h_{c2}$ are independent 
of the orientation of an applied magnetic field with respect to the 
crystallographic axes of these materials. 

\end{abstract}

\end{frontmatter}

\section{Introduction}

The Ginzburg-Landau (GL) theory predicts that, if the GL parameter $\kappa$ 
is temperature independent, the magnetic susceptibility $\chi$ of a 
type-II superconductor in the mixed state is a universal function 
of $h = H/H_{c2}(T)$ \cite{1}, i.e,   
\begin{equation}
 \chi(H,T)=\chi(h).
\end{equation}
As has recently been shown \cite{2}, this offers the possibility to evaluate 
the temperature dependence of the normalized upper critical field from 
measurements of the reversible isothermal magnetization in the mixed 
state without any additional assumptions. According to Ref. \cite{2}, the 
magnetizations of a sample in a fixed field $H$ at two different temperatures 
$T_{0}$ and $T$ are related by 
\begin{equation}
M(H,T_{0})=M(h_{c2}H,T)/h_{c2}+c_{0}(T)H,
\end{equation}
where $h_{c2}=H_{c2}(T)/H_{c2}(T_{0})$. While the first term on the right 
side of Eq. (2) follows straightforwardly from Eq. (1), the term 
$c_{0}(T)H$ accounts for the sample magnetization arising from any temperature 
dependent magnetic susceptibility of the superconductor in the normal state. 
By a suitable choice of the parameters $h_{c2}$ and $c_{0}(T)$, individual 
$M(H)$ curves measured at different temperatures may be merged 
into a single master curve $M(H,T_{0})$. In this way the temperature 
dependence of the normalized upper critical field $h_{c2}(T) = 
H_{c2}(T)/H_{c2}(T_{0})$ can be obtained \cite{2}. 

It is important to point out that the scaling procedure described by Eq. 
(2) provides values of the upper critical field as it enters the expression 
for the magnetization in the mixed state. This definition of $H_{c2}$ is 
unambiguous and the physical meaning of the upper critical field defined 
in this way is the same as in the GL theory. In the $H-T$ phase diagram 
the $H_{c2}(T)$ curve represents the upper boundary of the mixed state. 
This does not necessarily imply that in magnetic fields exceeding 
$H_{c2}(T)$, superconductivity in the sample is completely quenched. This 
applies particularly to high-$T_{c}$ superconductors (HTSC's) where, 
due to their rather specific properties, superconducting regions may exist 
in the form of separated islands or layers oriented along the direction 
of the magnetic field even in magnetic fields $H > H_{c2}(T)$. This is 
discussed in more detail in Ref. \cite{2}. In this case, there is no 
superconducting phase coherence in the direction perpendicular to the magnetic 
field and therefore, the magnetic flux is distributed between these islands 
or layers without the formation of a true mixed state. 

\section{Analysis of experimental data}

As in our previous work, we apply our scaling procedure to magnetization 
measurements available in the literature, but now concentrating on 
polycrystalline and ceramic samples. The samples characteristics are listed 
in Table I. Letters in the sample identification denote the chemical element 
characterizing the corresponding HTSC family. Because the $h_{c2}(T/Tc)$ curves 
for oxygen deficient YBa$_{2}$Cu$_{3}$O$_{7-x}$ (YBCO-123) and intrinsically 
underdoped Y$_{2}$Ba$_{4}$Cu$_{7}$O$_{15+x}$  (YBCO-247) samples are quite 
different from those of all other HTSC compounds, we present the results 
of our analysis in two different subsections. 

\subsection{Y-247 and oxygen deficient Y-123 materials}

Fig. 1(a) demonstrates the resulting $M(H)$ curves for $T_{0} = 44$ K, as 
obtained with our scaling procedure for two YBa$_{2}$Cu$_{3}$0$_{7-x}$ 
samples with $x = 0.19$ (Y$\#$2 and Y$\#$3). Although these two samples have 
the same oxygen content, their relative densities, with 90$\% $ for 
Y$\#$2 and 67$\%$ for Y$\#$3, and their average grain sizes are distinctly 
different [3]. Naturally this leads to noticeably different $M(H)$ curves, 
as may be seen in Fig. l(a). In spite of all these differences, the 
$h_{c2}(T)$ curves for these two samples are practically identical, as is 
demonstrated in the inset of Fig. 1(a). 
\begin{figure}[h]
 \begin{center}
  \epsfxsize=1\columnwidth \epsfbox {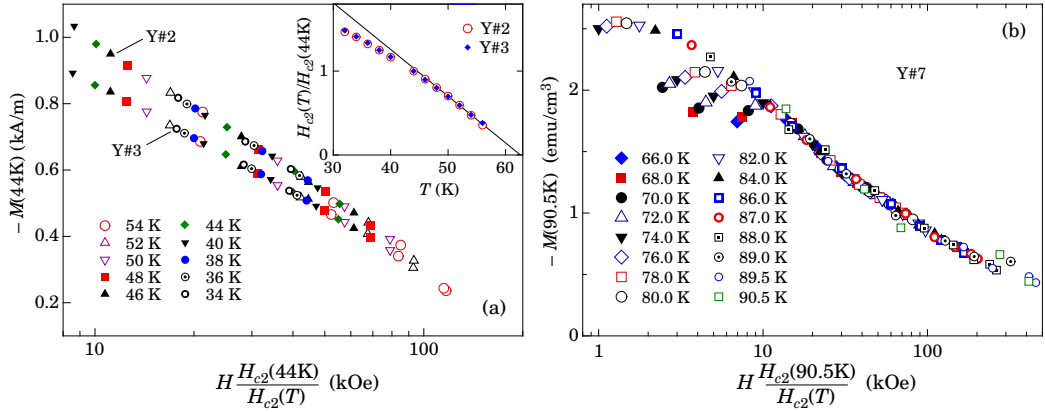}
  \caption{Magnetization data scaled according to Eq (2). (a) Samples 
           Y\#2 and Y\#3. The inset shows the corresponding $h_{c2}(T)$ 
           data. The solid line is the best linear fit to $h_{c2}(T)$ 
           for $T \ge 45$ K. (b) Sample Y\#7.}
 \end{center}
\end{figure}

The scaled $M(H)$ curve for an Y-247 sample (Y\#7) is shown in Fig. 
1(b). It is obvious that the scaling procedure is not valid at low 
temperatures and low magnetic fields. This failure is due to the fact that 
some of the original data points were measured below the irreversibility 
line and obviously do not correspond to the equilibrium 
magnetization.\footnote{It is often the case that the equilibrium 
magnetization below the irreversibility line is taken as the arithmetic 
mean of the values measured in increasing and decreasing fields, as it 
follows from the Bean critical state model \cite{bean1,bean2}. This model 
is based on the assumption that the critical current density $j_{c}$ is 
independent of the magnetic induction. However, it turns out that the values 
of $j_{c}$ for HTSC's, evaluated by application of the Bean critical-state 
model, depend on the applied magnetic field quite dramatically 
\cite{6,7,8,9,10,11,12,13,14}. Yet, for some mysterious reasons it is generally 
accepted that the Bean model is nevertheless applicable for this 
class of superconductors. Fig. 1(b) represents one additional evidence that 
the Bean model cannot provide a satisfactory quantitative description of 
the critical state in HTSC's.} Scaling is, however, well obeyed for 
$HH_{c2}(90.5$K$)/H_{c2}(T) \ge 10$ kOe. We present this example here in 
order to show that, if for some reasons the magnetic susceptibility does 
not follow Eq. (1), it is recognized by our procedure. 
\begin{figure}[h]
 \begin{center}
  \epsfxsize=0.52\columnwidth \epsfbox {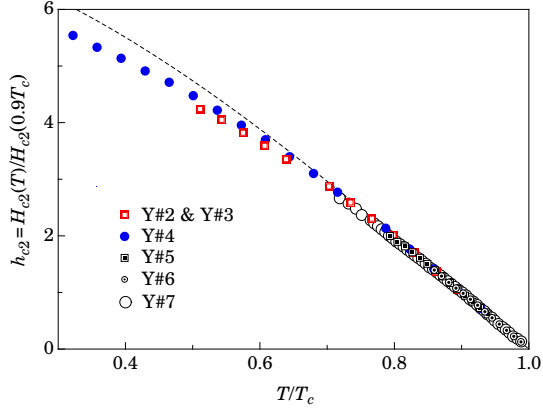}
  \caption{Temperature variations of $H_{c2}$, normalized by $H_{c2}(T = 
           0.9T_{c})$, for several oxygen deficient Y-123 and Y-247 
           samples. The dashed line represents a calculation based on the 
           weak coupling BCS theory \cite{gor,helf}}
 \end{center}
\end{figure}

As has been shown in Ref. \cite{2}, the $h_{c2}(T)$ curves for the oxygen 
deficient YBCO-123 samples are linear down to substantially lower 
temperatures than those for other HTSC's. This circumstance provides the 
possibility  of a reliable evaluation of $T_{c}$ by extrapolating the 
$h_{c2}(T)$ curve to $h_{c2} = 0$ even if the magnetization measurements 
are limited to temperatures not very close to $T_{c}$, as may be seen in 
the inset of Fig. 1(a). The $h_{c2}(T/T_{c})$ curves for several underdoped 
Y-123 samples, together with the results for two Y-247 samples, Y$\#$4 and 
Y$\#$5, are shown in Fig. 2. 

\subsection{$h_{c2}(T)$ for other HTSC compounds}

Fig. 3(a) shows the magnetization data for a Tl-based cuprate, sample 
Tl$\#$2, scaled according to Eq. (2). The mismatch between the scaled data 
points originating from $M(H)$ curves measured at different temperatures 
does not exceed the scatter of the original experimental data, which is 
obviously higher at higher temperatures. The resulting values of 
$h_{c2}(T) = H_{c2}(T)/H_{c2}(105$K$)$ for $T \ge 95$ K are plotted in 
Fig. 3(b). The data for $T \ge 107$ K, i.e., $T/T_{c} \ge 0.92$, lie along 
a straight line, while in a more extended temperature range, 
$h_{c2}(T)$ may quite well be approximated by a power law of the form
\begin{equation}
h_{c2}(T) = \frac{H_{c2}(T)}{H_{c2}(T_{0})} = \frac{1 - 
(T/T_{c})^{\mu}}{1 - (T_{0}/T_{c})^{\mu}}
\end{equation}
with two fit parameters $T_{c}$ and $\mu$. Because the magnetization was 
measured up to temperatures close to $T_{c}$, both the linear and the 
power-law extrapolation lead to practically the same value of $T_{c}$. 
In cases where the values of $h_{c2}$ sufficiently close to $T_{c}$ are 
not available, we employ Eq. (3) for the extrapolation of the $h_{c2}(T)$ 
curves to $h_{c2} = 0$. 
\begin{figure}[t]
 \begin{center}
  \epsfxsize=1\columnwidth \epsfbox {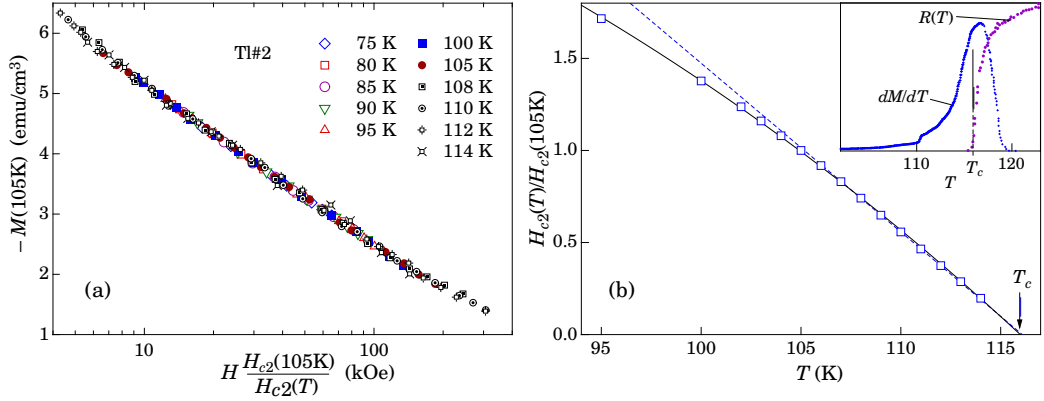}
  \caption{(a) Magnetization data for sample Tl\#2 scaled according to 
               Eq. (2) with $T_{0} = 105$ K. (b) 
               $H_{c2}(T)/H_{c2}(105$K) for the same sample. The solid 
               line is the best fit to $h_{c2}(T)$ with Eq. (3) for $T 
               \ge 95$ K. The dashed line is the best linear fit to 
               $h_{c2}(T)$ for $T \ge 107$ K. The inset shows the 
               temperature derivative of the low-field magnetization and 
               the zero-field resistance curves. The vertical line in the 
               inset indicates the position of $T_{c}$ evaluated by the 
               extrapolation of the $h_{c2}(T)$ curve.}
 \end{center}
\end{figure}

In the inset of Fig. 3(b) the vertical solid line indicates the position 
of $T_{c}$ evaluated with our scaling procedure with respect to experimental 
$dM/dT$ and $R(T)$ curves for the same sample. Note that $T_{c}$ does not 
coincide with any distinct point of these curves, such as the temperature 
where $R$ reaches zero or the maximum of the $dM/dT$ versus $T$ curve.
\begin{figure}[h]
 \begin{center}
  \epsfxsize=1\columnwidth \epsfbox {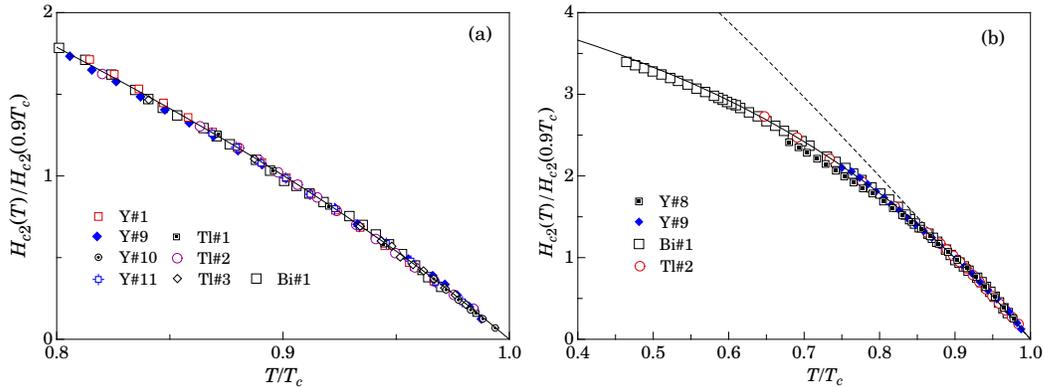}
  \caption{(a) and (b) Temperature variations of $H_{c2}$ normalized 
           by $H_{c2}(T = 0.9T_{c})$ for different samples listed 
           in Table I. The solid lines represent an analogous curve 
           drawn in Fig. 3(b) of Ref. \cite{2}. The dashed line represents 
           a calculation that is based to the weak coupling BCS theory 
           \cite{gor,helf}.}
 \end{center}
\end{figure}

In Fig. 4(a) we present $h_{c2}$ versus $T/T_{c}$ for several different 
polycrystalline and ceramic samples for $T \ge 0.8T_{c}$. Corresponding 
$h_{c2}(T/T_{c})$ data covering a more extended temperature range are shown 
in Fig. 4(b). The solid lines in Figs. 4(a) and 4(b) are simply transferred 
from Fig. 3(b) in Ref. \cite{2} and represent the $h_{c2}(T/T_{c})$ curve 
resulting from our analysis of magnetization measurements on several single 
crystals and grain-aligned Bi-based samples. It is obvious that the data 
points in Figs. 4(a) and 4(b) follow these solid lines rather closely. 
\begin{figure}[t]
 \begin{center}
  \epsfxsize=1\columnwidth \epsfbox {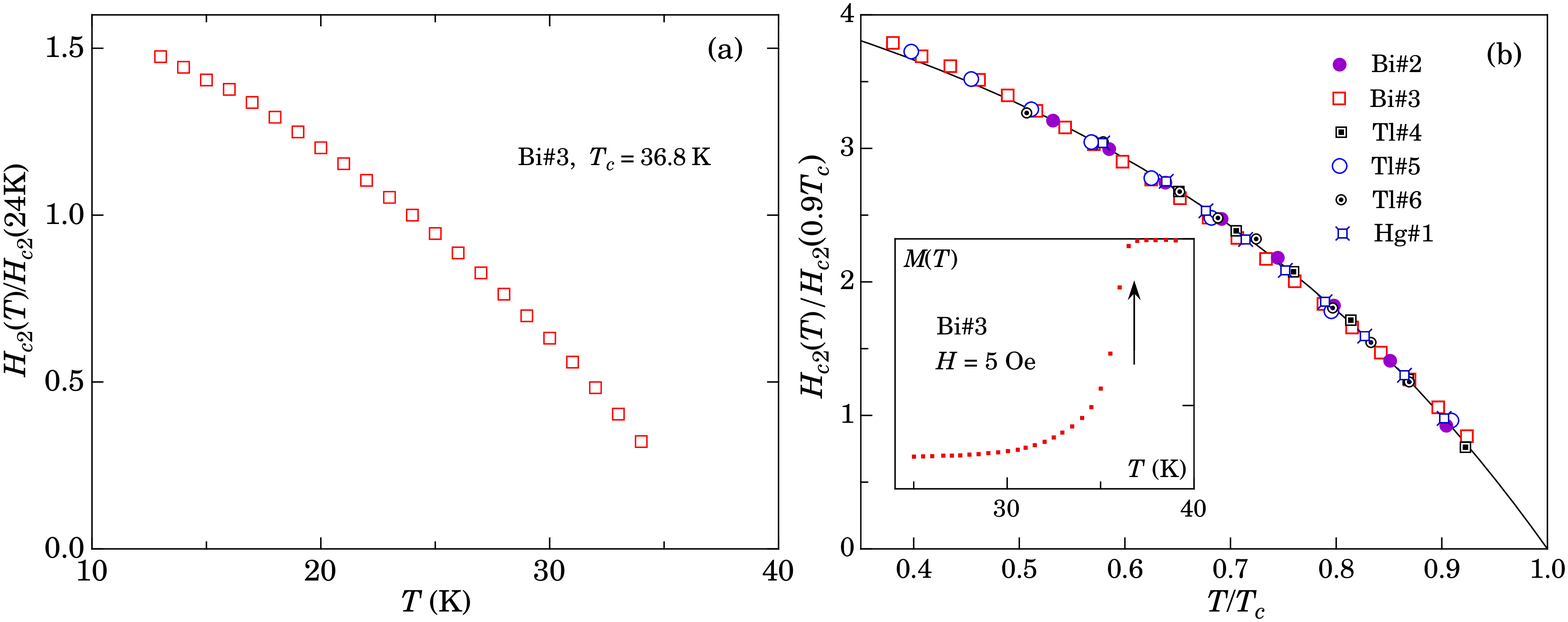}
  \caption{(a) $H_{c2}(T)/H_{c2}(24$K$)$ for a La-doped Bi-based sample 
           Bi$\#$3. (b) $H_{c2}(T)/H_{c2}(0.9T_{c})$ for several samples, 
           for which the values of $T_{c}$ were evaluated by fitting of 
           the $h_{c2}(T/T_{c})$ data to the solid line shown in Fig. 
           4(b). The inset shows the position of $T_{c}$ (vertical arrow) 
           estimated by this fitting procedure with respect to a low-field 
           magnetization curve for the Bi$\#3$ sample.}
 \end{center}
\end{figure}

For the procedure outlined above, as well as in our previous work 
\cite{2}, we have chosen only data from magnetization measurements that 
were made up to temperatures $T \sim 0.94-0.95T_{c}$, in order to obtain 
reliable values of $T_{c}$ by extrapolating the $h_{c2}(T)$ curves. A 
considerable number of studies available in the literature does not satisfy 
this criterion, however. As an example we show, in Fig. 5(a), the 
$h_{c2}(T)$ curve for a ceramic sample of 
Bi$_{1.95}$Sr$_{2.05-x}$La$_{x}$CuO$_{y}$ (Bi$\#$3). This curve is 
qualitatively very similar to those shown in Fig. 4 and in a next step 
we checked whether these and other analogous data may quantitatively be 
fit by the same solid lines as drawn in Figs. 4(a) and 4(b), simply by 
adjusting the value of $T_{c}$. The result of this procedure for several 
samples is shown in Fig. 5(b). Again all the $h_{c2}(T/T_{c})$ data follow 
the same solid line quite closely. The arrow in the inset of Fig. 6 indicates 
the position of $T_{c}$ determined in this way with respect to the 
field-cooled magnetization curve of sample Bi$\#$3. We applied this kind 
of analysis also to a number of magnetization studies of single crystals 
and grain-aligned samples 
\cite{22,23,24,25,26,27,28,29,30,31,32,33,34,35,36,37}. Again in all these 
cases, the $h_{c2}(T)$ curves may very well be fitted to the universal 
$h_{c2}(T/T_{c})$ curve shown in Fig. 4(b) and originally displayed 
in Fig. 3(b) of Ref. \cite{2}.

\section{Conclusion}

It turns out that the scaling procedure described by Eq. (2) works quite 
well not only for single crystals or grained-aligned samples of HTSC's, 
as previously demonstrated in Ref. \cite{2}, but also for HTSC polycrystalls 
and ceramics. Because polycrystalline samples can be made much larger than 
single crystals, the magnetization measurements can, in principle, be 
made more accurately and may easily be extended up to temperatures very 
close to $T_{c}$. The fact that our scaling procedure works for polycrystalline 
or ceramic samples means that the shape of the normalized $H_{c2}(T)$ curves 
does not depend on the orientation of the external magnetic field with 
respect to the crystallographic axes. This is undoubtedly the most important 
result of this study. 

The results of our analysis presented in this work confirm the main result 
of our previous study where it was shown that the normalized temperature 
variations of the upper critical field follow the same universal curve 
for most of the HTSC's \cite{2}. It now appears that all HTSC's may be divided 
into two groups with two distinctly different normalized $H_{c2}(T)$ curves 
(see Figs. 2 and 4). For the smaller group consisting of Y-247 and oxygen 
deficient Y-123 compounds the temperature dependence of $H_{c2}$ is quite close 
to that calculated on the basis of the weak-coupling BCS theory 
\cite{gor,helf}. For the other group, which includes practically all other 
HTSC compounds, all the normalized $h_{c2}(T/T_{c})$ curves are identical but 
are well below the calculated values of Refs. \cite{gor,helf}. At present, 
it is not clear what causes the difference between the two sets of 
$h_{c2}(T/T_{c})$ curves. We may only state that $h_{c2}(T/T_{c})$ 
of a YBa$_{2}$Cu$_{3}$O$_{6.85}$ sample is already completely different 
from that of optimally doped Y-123 material. 
\newpage
\begin{table}[h]
\caption{Sample identification.}
\begin{tabular}{lcccccc}
\multicolumn{1}{c}{No.} &
\multicolumn{1}{c}{Refs.} &
\multicolumn{1}{c}{Compound} &
\multicolumn{1}{c}{$T_{c}$ (K)} \\

 Y\# 1 & \cite{3} & YBa$_2$Cu$_3$O$_{6.98}$ & 92.1 \\
 Y\# 2 & \cite{3} & YBa$_2$Cu$_3$O$_{6.81}$ & 62.0 \\
 Y\# 3 & \cite{3} & YBa$_2$Cu$_3$O$_{6.81}$ & 62.0 \\
 Y\# 4 & \cite{3} & YBa$_2$Cu$_3$O$_{6.69}$ & 55.9 \\
 Y\# 5 & \cite{Y5} & YBa$_2$Cu$_3$O$_{6.5}$ & 44.8 \\
 Y\# 6 & \cite{13} & Y$_2$Ba$_4$Cu$_7$O$_{15+x}$ & 93.1 \\
 Y\# 7 & \cite{trisc} & Y$_2$Ba$_4$Cu$_7$O$_{15+x}$ & 91.6 \\
 Y\# 8 & \cite{trisc} & YBa$_2$Cu$_4$O$_{8}$ & 80.7 \\
 Y\# 9 & \cite{trisc} & YBa$_2$Cu$_3$O$_{7-x}$ & 92.9 \\
 Y\#10 & \cite{Y10} & YBa$_2$Cu$_3$O$_{7-x}$ & 92.1 \\
 Y\#11 & \cite{Y11} & YBa$_2$Cu$_3$O$_{7-x}$ & 92.1 \\
Bi\# 1 & \cite{Bi1} & Bi$_{2.12}$Sr$_{1.9}$Ca$_{1.2}$Cu$_{1.96}$O$_{8+x}$ & 86.1 \\
Bi\# 2 & \cite{trisc} & Bi$_2$Sr$_2$CaCu$_2$O$_{8+x}$ & 93.2 \\
Bi\# 3 & \cite{Bi3} & Bi$_{1.95}$Sr$_{2.05-x}$La$_{x}$CuO$_{y}$ & 36.8 \\
Tl\# 1 & \cite{Tl1} & Tl$_2$Ba$_2$Ca$_2$Cu$_3$O$_{10+x}$ &122.8 \\
Tl\# 2 & \cite{Tl2} & 
Tl$_{0.7}$Pb$_{0.2}$Bi$_{0.2}$Sr$_{1.8}$Ba$_{0.2}$Ca$_{1.9}$Cu$_3$O$_{10+x}$ &122.8 \\
Tl\# 3 & \cite{trisc} & Tl$_2$Ba$_2$CaCu$_2$O$_{8+x}$ &107.1 \\
Tl\# 4 & \cite{trisc} & Tl$_2$Ba$_2$Cu$_2$O$_{6+x}$ & 92.1 \\
Tl\# 5 & \cite{trisc} & Tl$_2$Ba$_2$CaCu$_2$O$_{x}$ & 90.0 \\
Tl\# 6 & \cite{6} & Tl$_{0.5}$Pb$_{0.5}$Sr$_4$Cu$_2$CO$_3$O$_{7}$ & 69 \\
Hg\# 1 & \cite{6} & Hg$_2$Ba$_2$Ca$_2$Cu$_3$O$_{x}$ &133.0 \\
\end{tabular}
\end{table}
%


\end{document}